\documentclass[aps,prb,twocolumn,groupedaddress,showpacs,final,reprint]{revtex4-1}
\usepackage{graphicx}
\usepackage{color}

\begin{document}


\title{Thermodynamics of the Bose-Hubbard model in a Bogoliubov+U theory }


\author{Dario H{\"u}gel}
\author{Lode Pollet}
\affiliation{Department of Physics, Arnold Sommerfeld Center for Theoretical Physics and Center for NanoScience, Ludwig-Maximilians-Universit{\"a}t M{\"u}nchen, Theresienstrasse 37, 80333 Munich, Germany}


\date{\today}

\begin{abstract}
We derive the Bogoliubov+U formalism to study the thermodynamical properties of the Bose-Hubbard model. The framework can be viewed as the zero-frequency limit of bosonic dynamical mean-field theory (B-DMFT), but equally well as an extension of the mean-field decoupling approximation in which pair creation and annihilation of depleted particles is taken into account. The self-energy on the impurity site is treated variationally, minimizing the grand potential. The theory containing just three parameters that are determined self-consistently reproduces the  $T=0$ phase diagrams of the three-dimensional and  two-dimensional Bose-Hubbard model  with an accuracy of $1 \%$ or better. The superfluid to normal transition at finite temperature is also reproduced well and only slightly less accurately than in B-DMFT.
\end{abstract}


\maketitle
\section{Introduction}
The properties of cold atomic gases trapped in an optical lattice can be tuned and controlled very precisely, providing a powerful tool for the simulation of the iconical low-energy effective Hamiltonians of condensed-matter models [\onlinecite{BlochRev}]. Dramatic experimental progress in this field, such as the observation of the Mott insulator to superfluid phase transition in the Bose-Hubbard model [\onlinecite{BlochRev}] or the recent realization of the Hofstadter model [\onlinecite{Hof}], have galvanized the condensed matter community. 

The accuracy of cold atom experiments has laid bare the need for advanced numerical methods in order to tackle these correlated many-body systems quantitatively. In one dimension the density matrix renormalization group (DMRG) [\onlinecite{Kuhner, Rapsch, Kollath1, Kollath2}] works very well, while in higher dimensions the numerical complexity represents a problem. Very successful also in higher dimensions have been path integral quantum Monte Carlo (QMC) simulations with worm-type updates [\onlinecite{Tupitsyn}] leading to identical results for the Bose-Hubbard model as observed in experiment [\onlinecite{Trotzky},\onlinecite{Rev_Lode}].
Despite all its impressive successes for bosonic systems the worm-algorithm suffers from a prohibitive sign problem when cold atoms are subject to (artifical) gauge fields. In such cases no algorithm is known that works and one is forced to resort to approximations. This has been a main driving force for the development of bosonic dynamical mean-field theory [\onlinecite{BDMFT},\onlinecite{BDMFT2}] (B-DMFT).

In B-DMFT, as in standard mean-field theory, the many-body system is projected onto a single impurity, keeping only the local propagators. This provides us with a model in which the self-energy and the local propagators can be determined self-consistently by solving an effective impurity action and a self-consistency condition iteratively.
At the moment it has only been formulated for single-site impurities, but the ultimate goal is to formulate it for small clusters that can also deal with larger unit cells.
It is known that B-DMFT provides excellent agreement [of the order of $1 \%$ in three dimensions (3D)] with experimental and QMC data [\onlinecite{BDMFT}] for the standard Bose-Hubbard model and improves remarkably on static mean-field theory. B-DMFT is hence a promising candidate to deal with more complicated dispersions: The hope is that it deals with the local interactions as accurately as in the standard Bose-Hubbard model whereas a small cluster could treat the full dispersion. Furthermore, one would expect that a cluster method goes beyond real-space DMFT for multi-component systems and systems with long-range interactions and/or dispersions [\onlinecite{Hofst1},\onlinecite{Hofst3}].

However, B-DMFT is numerically rather complex due to the imaginary-time dependency of the hybridization term. At finite temperature the impurity problem has to be solved by continuous time Monte Carlo methods [\onlinecite{BDMFT},\onlinecite{BDMFT2}], where, due to the difference in sign between the normal and the anomalous Green's function, a sign problem arises in the symmetry broken phase.

In this paper, we filter out the ingredients of B-DMFT that are indispensable for its accuracy and arrive at a simpler formalism. This is the  Bogoliubov+U theory (B+U), which makes use of a simplified effective impurity Hamiltonian, similar to the action of extended mean field theory, which was recently developed in the high-energy community [\onlinecite{EMFT1},\onlinecite{EMFT2}] but differs conceptually from our formalism. B+U has a negligible computational cost and is not prone to numerical instabilities. The premise of our theory is that the Bose-Hubbard model can be fully characterized at zero temperature by the three parameters $\phi$, $\Sigma_{11}$, and $\Sigma_{12}$ (the condensate order parameter, and the normal and the anomalous self-energy, respectively) if the self-energy is treated as a variational parameter, providing a far better approximation to finite-temperature properties than simple mean-field theory.
B+U can be seen as a simplified B-DMFT where only a single Matsubara frequency is kept (and is hence conserving). It is different from the variational cluster approximation (VCA) by also considering non zero values of pair creation and annihilation of depleted particles [\onlinecite{VCA}]. It is also the simplest accurate extension of the weakly interacting Bose gas theory  [\onlinecite{Svist-Gior-Poll}] to lattice systems with a superfluid to Mott insulator transition. It further provides a very natural framework compared to the collective quantum field theory developed by Kleinert {\it et al.} [\onlinecite{Kleinert}] and the two collective fields proposed by Cooper {\it et al.} [\onlinecite{Cooper1,Cooper2,Cooper3}], and behaves quantitatively much better. It has a similar functional degree of freedom as the projector technique introduced by Trefzger and Sengupta [\onlinecite{Trefzger}] for finite lattices. Among other methods which are used to treat the Bose-Hubbard model are also the process chain approach [\onlinecite{Eckardt},\onlinecite{Teichmann}] and the quantum phase field $U(1)$ rotor field [\onlinecite{Polak}]. 

The paper is organized as follows. In Sec. \ref{sec2} the B+U formalism is derived for the Bose-Hubbard model in equilibrium, while in Sec. \ref{sec3} the details of the variational calculation of the optimal self-energy are shown. We furthermore summarize the full self-consistent scheme of B+U and show how thermodynamic quantities can be calculated from it in Sec. \ref{sec4}, while some simple limits of B+U are explained in Sec. \ref{sec41}. In Sec. \ref{sec5} we present the results at zero and finite temperature comparing them with QMC and B-DMFT. Finally, in Sec. \ref{sec6} we conclude and present a short outlook about future applications of the B+U formalism.
\section{Solver and self-consistency condition}
\label{sec2}
In this section we derive the B+U formalism for the Bose-Hubbard model in equilibrium. In order to derive an effective Hamiltonian, we start from the full Bose-Hubbard Hamiltonian,
\begin{equation}
\label{formela}
H_{\rm BH} =-J\sum_{\langle{i,j}\rangle}b^{\dagger}_{i}b^{}_{j}+\frac{U}{2}\sum_{i}n^{}_{i}\left(n^{}_{i}-1\right)-\mu\sum_{i}n^{}_{i},
\end{equation}
where $b^{\dagger}_{i}$ is a bosonic single-particle creation operator on lattice-site $i$, $n^{}_{i}$ is the particle-number operator, $J$ denotes the tunneling amplitude, $U$ the on-site interaction, $\mu$ the chemical potential, and  $\langle{i,j}\rangle$ means that we sum over nearest neighbors. 
In order to determine the thermodynamic properties of the system, we have to compute the condensate density and the connected Green's function, defined respectively as
\begin{eqnarray}
\phi & = & \left < b \right>, \\
{\bf G}_{c}^{i,j}(\tau) & = & -\left< \Phi_i(\tau) \Phi_j^{\dagger}(0) \right> + \phi \phi^{\dagger},
\label{formelg}
\end{eqnarray}
with Nambu notation  $\Phi_{i}(\tau)=\left(  \begin{array}{c} b_{i}(\tau) \\  b^{\dagger}_{i}(\tau) \end{array} \right) $. The possibility of broken $U(1)$ symmetry forces $b_{j}$ to be expanded around its mean-field value $\phi=\left<  b_{j} \right>$ (which we take to be site independent and can always be chosen real) by $b_{j}=\phi+\delta b_{j}$. If we concentrate on the site at the origin $b_{o}$, the Hamiltonian can be rewritten as
\begin{eqnarray}
H & = & H_o + H_{\rm ext} + \Delta H, \nonumber \\
H_o & = & \frac{U}{2}  n_o(n_o- 1) - \mu n_o -zJ \phi\left(b_o+b^{\dagger}_o \right), \\ 
\Delta H & = &  - J\sum_{\langle{i,o}\rangle}\left(\delta b^{\dagger}_{i}\delta b_{o}+\delta b^{\dagger}_{o}\delta b_{i}\right),  \nonumber
\end{eqnarray}
where $H_{\rm ext}$ contains all terms of the Hamiltonian (\ref{formela}) not containing the origin ``$o$''.
The notation $\langle{i,o}\rangle$ means that we sum over the nearest neighbors of $o$, and $z$ is the coordination number. 
We wish to separate the full partition function $Z={\rm tr}\left[e^{-\beta H}\right]$ as $Z = Z_{\rm ext} Z_o$. Here, $Z_{\rm ext}$ is the full (and unknown) partition of the system determined by the terms in the Hamiltonian not involving the site $o$. It is treated as an irrelevant number in the rest of the paper. The partition function $Z_o$ contains the full local Hamiltonian $H_o$ as well as the correlations introduced by ``ext'' on the origin as follows,
\begin{equation}
Z_o =  {\rm tr}\left[e^{- \beta\left(H_{o}+\left< \Delta H \right>_{H_{\rm ext}} \right)}  \right].
\end{equation}
We approximate the expectation value $\langle\Delta H\rangle_{H_{\rm ext}}$ by the cumulant expansion
\begin{eqnarray}
\label{formele}
\left< \Delta H \right>_{H_{\rm ext}} & \approx &  -J\left\langle\sum_{\langle{i,o}\rangle}\delta \Phi^{\dagger}_{i}\delta \Phi_{o}\right\rangle_{H_{\rm ext}} \nonumber \\
& & -\frac{1}{2}J^{2}\left\langle\sum_{\langle{i,o}\rangle}\delta\Phi^{\dagger}_{i}\delta\Phi_{o}\sum_{\langle{j,o}\rangle}\delta\Phi^{\dagger}_{j}\delta\Phi_{o}\right\rangle_{H_{\rm ext}} \nonumber \\
& = & 0-\frac{1}{2}\delta \Phi^{\dagger}_{o}{\bf{\Delta}}\delta\Phi_{o},
\end{eqnarray}
where $\langle \Delta H\rangle$ and $\langle \Delta H  \Delta H\rangle$ are rewritten in terms of Nambu operators and  ${\bf{\Delta}}$ is an unknown $2\times 2$ real-valued matrix with two independent components $\Delta_{11}=\Delta_{22}$ and $\Delta_{12}=\Delta_{21}$ which describes a correction to the common mean-field impurity Hamiltonian. The anomalous term $\Delta_{12}$, containing processes of the type $\delta b^{2}$, is explicitly taken to be finite in this notation, since it is known from the Bogoliubov theory that deep in the superfluid phase it becomes equally important to the normal (diagonal) term $\Delta_{11}$, containing the $\delta b^{\dagger}\delta b$ terms.  By (\ref{formele}) we arrive at the effective impurity Hamiltonian
\begin{eqnarray}
H_E & = & - \frac{1}{2} \delta\Phi^{\dagger}_o {\bf{\Delta}} \delta\Phi_o -zJ\phi\left(b_o+b^{\dagger}_o \right) \nonumber \\
{} & {} & +  \frac{U}{2}  n_o (n_o - 1) - \mu n_o.  \label{formelf}
\end{eqnarray}
As can be seen in the Appendix, this effective impurity Hamiltonian is equivalent to B-DMFT in the limit
\begin{equation}
{\bf \Delta}(\tau_1 - \tau_2) \to {\bf \Delta} \delta(\tau_1 - \tau_2).
\label{eq:hyb}
\end{equation}
Since $\bf\Delta$ is independent on imaginary time, the Dyson equation on the impurity has to be evaluated only for a single Matsubara value. The Green's function which mirrors the symmetry relations assumed for $\bf\Delta$ is the one evaluated at $\omega_n =0$.
The central characteristic of B+U theory is that we demand that the condensate and the (full) Green's function of the Bose-Hubbard model evaluated at o {\it for zero (Matsubara) frequency} coincide with the one of system (\ref{formelf}), i.e.,
\begin{eqnarray}
\phi & \equiv & \left< b_o \right>_{H_E},  \label{eq:selfcon1} \\
{\bf G}_{c}^{o,o}(\omega_n = 0) & \equiv & -\left<( \Phi_{o} \Phi^{\dagger}_{o})(\omega_n = 0) \right>_{H_{E}} + \phi\phi^{\dagger}. \label{eq:selfcon2}
\end{eqnarray}
The paradoxical compatibility of Eqs.(\ref{eq:selfcon2}) and (\ref{eq:hyb}) is specific for bosonic systems [see also below Eq. (\ref{formelh})].
The equations constitute a self-consistency problem, whose solution also fixes the factors $ \Delta_{11}$ and $ \Delta_{12}$.
This can be solved in a unique way if ${\bf \Delta}[ {\bf G}_c, \phi]$ is invertible, or, technically speaking, if the Luttinger-Ward functional is unique. Since the static mean-field (i.e., the decoupling approximation with $\Delta_{11} = \Delta_{12}=0$) is always a solution, it is easy to convince oneself that multiple (local) minima occur (cf. Ref. [\onlinecite{Kozik}] for a recent discussion). Nevertheless, we have been able to determine the physically correct solution without problem in all parameter regimes (see below).
In practice, one uses an iteration scheme to solve the self-consistency problem. The factors $\Delta_{11}$ and $ \Delta_{12}$ follow from the Dyson equation on the impurity site at zero Matsubara frequency,
\begin{equation}
\label{formelh}
{\bf{\Delta}}={\bf{\Sigma}_{E}}(\omega_n=0) + {\bf{G}_{c,E}}(\omega_n=0)^{-1}-\mu{\bf{I}},
\end{equation}
with an unknown self-energy $\bf{\Sigma}_{E}$. The connected Green's function on the impurity site ${\bf{G}_{c,E}}$ is calculated through the full Green's function on the lattice by 
\begin{equation}
\label{formeli}
{\bf{G}_{c,E}}(\omega_n=0)=\frac{1}{(2\pi)^{d}}\int{d^{d}k}{\bf{G}}(\omega_n = 0,k),
\end{equation}
and the Dyson equation of the full lattice
\begin{equation}
\label{formelj}
{\bf{G}}(\omega_n = 0,k)^{-1}={\bf{G}_{0}}(\omega_n = 0,k)^{-1}-{\bf{\Sigma}_{E}} (\omega_n=0),
\end{equation}
with the bare Green's function given by ${\bf{G}_{0}}(\omega_{n},k)^{-1}=\left[\mu-\epsilon(k)\right]{\bf{I}}+i\omega_{n}{\bf{\sigma}_{z}}$, where $\epsilon(k)$ is the dispersion relation of the lattice and $\omega_{n}=\frac{2\pi}{\beta}n$ are the Matsubara frequencies. 

The approximation (\ref{eq:hyb}) shows that the B+U theory has the same functional form as the decoupling approximation in the non broken phase. In that case only $\Delta_{11}$ is present but it acts as a shift in chemical potential. For the broken phase, $\Delta_{11}$ and $\mu$ are combined with different operators. They control the density of the condensed and depleted atoms, whereas $\Delta_{12}$ mainly determines the anomalous density. According to the Bogoliubov theory of the weakly interacting Bose gas, the anomalous propagator is equally important (but opposite in sign) as the normal propagator deep in the superfluid phase. In this way, the deep superfluid regime is taken care of appropriately in our formalism. The Mott localization is enabled by the exact treatment of the density fluctuations on the impurity.

\section{Variation of the Self-Energy}
\label{sec3}
In order to solve the impurity problem, we consider the minimization of its grand potential $\Omega[{\bf{\Sigma}},\phi]$ with respect to its self-energy ${\bf{\Sigma}}$, as is also done in self-energy-functional theory [\onlinecite{SFT}] and VCA [\onlinecite{VCA}]. The minimum with respect to the kinetic condensate term $zJ\phi$, $\frac{\delta\Omega}{\delta(zJ\phi)}=\frac{\delta\Omega}{\delta(zJ\phi^{*})}=0$, is already taken care of by the self-consistency condition (\ref{eq:selfcon1}). $\phi$ is thus kept constant during the variational calculation of the self-energy. We therefore have to minimze
\begin{equation}
\label{formelk}
\frac{\delta\Omega}{\delta{\bf\Sigma}}=\frac{\delta\Omega}{\delta\Delta_{11}}\frac{\delta\Delta_{11}}{\delta{\bf\Sigma}}+\frac{\delta\Omega}{\delta\Delta_{12}}\frac{\delta\Delta_{12}}{\delta{\bf\Sigma}}.
\end{equation}
We are able to find an analytic expression of $\frac{\delta\Omega}{\delta\Delta_{ij}}$, since the grand potential is defined as $\Omega({\bf\Delta},\phi)=-\text{ln}Z({\bf\Delta},\phi)$ with $Z({\bf\Delta},\phi)=\text{Tr}[e^{-\beta{H_{E}({\bf\Delta},\phi)}}]$, giving us
\begin{eqnarray}
\frac{\delta\Omega({\bf\Delta},\phi)}{\delta\Delta_{11}} & = & \langle{2\phi b_o-n_o}\rangle_{H_{E}({\bf\Delta},\phi)}-|\phi|^{2}, \\
\frac{\delta\Omega({\bf\Delta},\phi)}{\delta\Delta_{12}} & = & \langle{2\phi b_o-b_{o}^{2}}\rangle_{H_{E}({\bf\Delta},\phi)}-|\phi|^{2}.
\label{formell}
\end{eqnarray}
After integration, this gives us the relation
\begin{equation}
\label{formelm}
\Omega({\bf{\Sigma}},\phi)=A({\bf \Delta}({\bf{\Sigma}}),\phi)+B({\bf\Delta}({\bf{\Sigma}}),\phi)+C.
\end{equation}
with some irrelevant integration constant $C$ and
\begin{eqnarray}
A({\bf\Delta}({\bf\Sigma}),\phi) & = & \frac{\delta\Omega({\bf\Delta}({\bf\Sigma}),\phi)}{\delta\Delta_{11}}\Delta_{11}({\bf\Sigma}), \\
B({\bf\Delta}({\bf\Sigma}),\phi) & = & \frac{\delta\Omega({\bf\Delta}({\bf\Sigma}),\phi)}{\delta\Delta_{12}}\Delta_{12}({\bf\Sigma}).
\label{formeln}
\end{eqnarray}
In order to avoid unphysical results, we have to introduce upper bounds on $|\Delta_{ij}|$. From (\ref{formele}) we see that ${\bf\Delta}$ cannot exceed the kinetic energy of a double hopping process of depleted particles,
\begin{equation}
\label{formelo}
|\Delta_{12}|\leq|\Delta_{11}|\leq(zJ)^{2}\langle{\delta b^{\dagger}_o \delta b^{}_o}\rangle.
\end{equation}
Furthermore, we require that for all momenta $k$  $G_{11}(\omega_{n}=0,k)^{-1}\leq{-\epsilon}$ and $\text{det}[{\bf{G}}(\omega_{n}=0,k)^{-1}]\geq\epsilon$ [where a small $\epsilon$ is introduced for stability requirements when inverting the $2$x$2$ matrix ${\bf{G}}(\omega_{n}=0,k)^{-1}$ in (\ref{formelj})], giving us additional bounds on the self-energy
\begin{eqnarray}
\Sigma_{11} & \geq & \mu+zJ+\epsilon, \label{formelp1}\\
|\Sigma_{12}| & \leq & \sqrt{(\Sigma_{11}-zJ-\mu)^{2}-\epsilon^{2}}. \label{formelp2}
\end{eqnarray}
\section{Full scheme and Observables}
\label{sec4}
By combining Secs. \ref{sec2} and \ref{sec3} we can write down the full self-consistent scheme for the B+U theory. Starting from an initial guess for $\phi$ and ${\bf{\Delta}}$ (usually the converged mean-field values for ${\bf{\Delta}}={\bf 0}$), we calculate a new value for $\phi$ through Eqs. (\ref{formelf}) and (\ref{eq:selfcon1}). Then we search for the optimal value of the self-energy by minimizing (\ref{formelm}) while keeping $\phi$ constant. This is done by varying $\bf{\Sigma}$ within the bounds (\ref{formelp1}) and (\ref{formelp2}) and calculating ${\bf\Delta}({\bf\Sigma})$ by Eqs. (\ref{formelh})-(\ref{formelj}), keeping in mind the bound on ${\bf\Delta}$ (\ref{formelo}). Once the optimal value $\bf\Sigma_{opt}$ is found, the new value for ${\bf\Delta}$, ${\bf\Delta}({\bf\Sigma_{opt}})$, is plugged into (\ref{formelf}), from which a new value for $\phi$ is calculated. This procedure is repeated until convergence is reached.
In the B+U self-consistency all bonds adjacent to $o$ are included in $H_E$, whereas when computing the quantities per site, all bonds have to be counted only once. In order to calculate the correct thermodynamic quantities per site once convergence is reached, one therefore has to divide $\bf\Delta$ by 2, giving us e.g. for the density per site $\langle n \rangle=N/V$,
\begin{equation}
\label{formelr3}
\langle n \rangle\rightarrow\langle n_o \rangle_{H_E ({\bf\Delta}/2,\phi)}.
\end{equation}
We can further divide the Hamiltonian into a kinetic [upper line in (\ref{formelf})] and a potential term [lower line in (\ref{formelf})], giving us expressions for the kinetic and potential energy  per site
\begin{eqnarray}
E_{\rm kin} & = & -\frac{1}{2}\left(\Delta_{11} \langle \delta b^{\dagger}_{o}\delta b^{}_{o} \rangle+\Delta_{12}\langle \delta b_{o}^{2} \rangle\right)-zJ |\phi|^{2} , \\
E_{\rm pot} & = & \frac{U}{2}\left(\langle n_{o}^{2} \rangle - \langle n_{o} \rangle\right) - \mu  \langle n_{o} \rangle,
 \label{formels}
\end{eqnarray}
where the total energy per site is given by $E_{\rm tot}=E_{\rm kin}+E_{\rm pot}$.
It should further be noted that even though we do not need to calculate ${\bf G}(\tau)$ explicitly in the solver and we do not include any retardation in our formalism, we can still calculate correlation functions of the kind $\langle A(\tau)B(0) \rangle$ by
\begin{equation}
\langle{A(\tau)B(0)}\rangle_{H_E}=\frac{1}{Z}{\rm tr}\left[e^{-(\beta-\tau) H_{E}}Ae^{-\tau H_{E}}B\right],
\label{formelt}
\end{equation}
or directly in energy space through the eigenvalues of $H_E$. Since the B+U solver consists of a single impurity, in order to compute momentum-dependent quantities, one has to resort to the approximate expression
\begin{equation}
\label{formels1}
{\bf{G}}(i\omega_n,k)^{-1}={\bf{G}_{0}}(i\omega_n,k)^{-1}-{\bf{\Sigma}} (i\omega_n),
\end{equation}
with
\begin{equation}
{\bf \Sigma}(i\omega_n)=i\omega_n  \sigma_z+\mu{\bf I}+ {\bf \Delta}-{\bf G_{c,E}}^{-1}(i\omega_n).
\label{formels2}
\end{equation}
By a Fourier transformation this enables us to compute such quantities as the momentum-dependent density
\begin{equation}
n(k)=-{\bf{G}}(k,\tau=0^{+})-1,
\label{formels3}
\end{equation}
or the critical quasi particle and quasi hole energies at zero momentum $\epsilon_{p/h}$ which can be evaluated from the asymptotic behavior of ${\bf{G}}(k=0,\tau)$ at zero temperature through [\onlinecite{QMC_Cap},\onlinecite{Duchon}]
\begin{equation}
{\bf{G}}(k=0,\tau)\rightarrow \left \lbrace \begin{array}{ll} Z_p e^{\epsilon_p \tau}\text{      } & \tau\rightarrow + \infty, \\  
Z_m e^{-\epsilon_m \tau}\text{      } & \tau\rightarrow - \infty,
 \end{array}\right.
\label{formels4}
\end{equation}
where $Z_p = Z_m-1$.

\section{Simple Limits}
\label{sec41}
From the relation (\ref{formelo}) it is clear that ${\bf{\Delta}}\rightarrow{\bf 0}$ as $J\rightarrow 0$. Furthermore, also, as $U$ goes to zero, ${\bf{\Delta}}$ vanishes, since $\langle{\delta b^{\dagger}_o \delta b^{}_o}\rangle \rightarrow 0$. Therefore in both cases the mean-field limit is recovered. In the case of $U\ll J$ the mean-field limit is consistent with the weakly interacting Bose gas theory [\onlinecite{Svist-Gior-Poll}], where the self-energy is frequency independent as is the case for B+U in our approach.
Another simple limit of B+U is the Bethe lattice for a semicircular density of states given by
\begin{equation}
D(\epsilon)=\frac{1}{2\pi z J^{2}}\sqrt{4zJ^{2}-\epsilon^{2}},\text{ }|\epsilon|\leq 2\sqrt{z}J, 
\label{formelt1}
\end{equation}
as was also implemented for B-DMFT [\onlinecite{BDMFT},\onlinecite{BDMFT2}] , which reduces the self-consistency of B+U to one single equation,
\begin{equation}
{\bf\Delta}=-zJ^{2}{\bf G_{c,E}}(\omega=0).
\label{formelt2}
\end{equation}
As can be seen in Sec. \ref{sec5} for the 3D case this leads to good agreement with the full self-consistency of a cubic lattice with a much lower numerical cost, since the minimization routine described in Sec. \ref{sec3} is no longer necessary.
\section{Results}
\label{sec5}
\begin{figure}
\includegraphics[scale=.35]{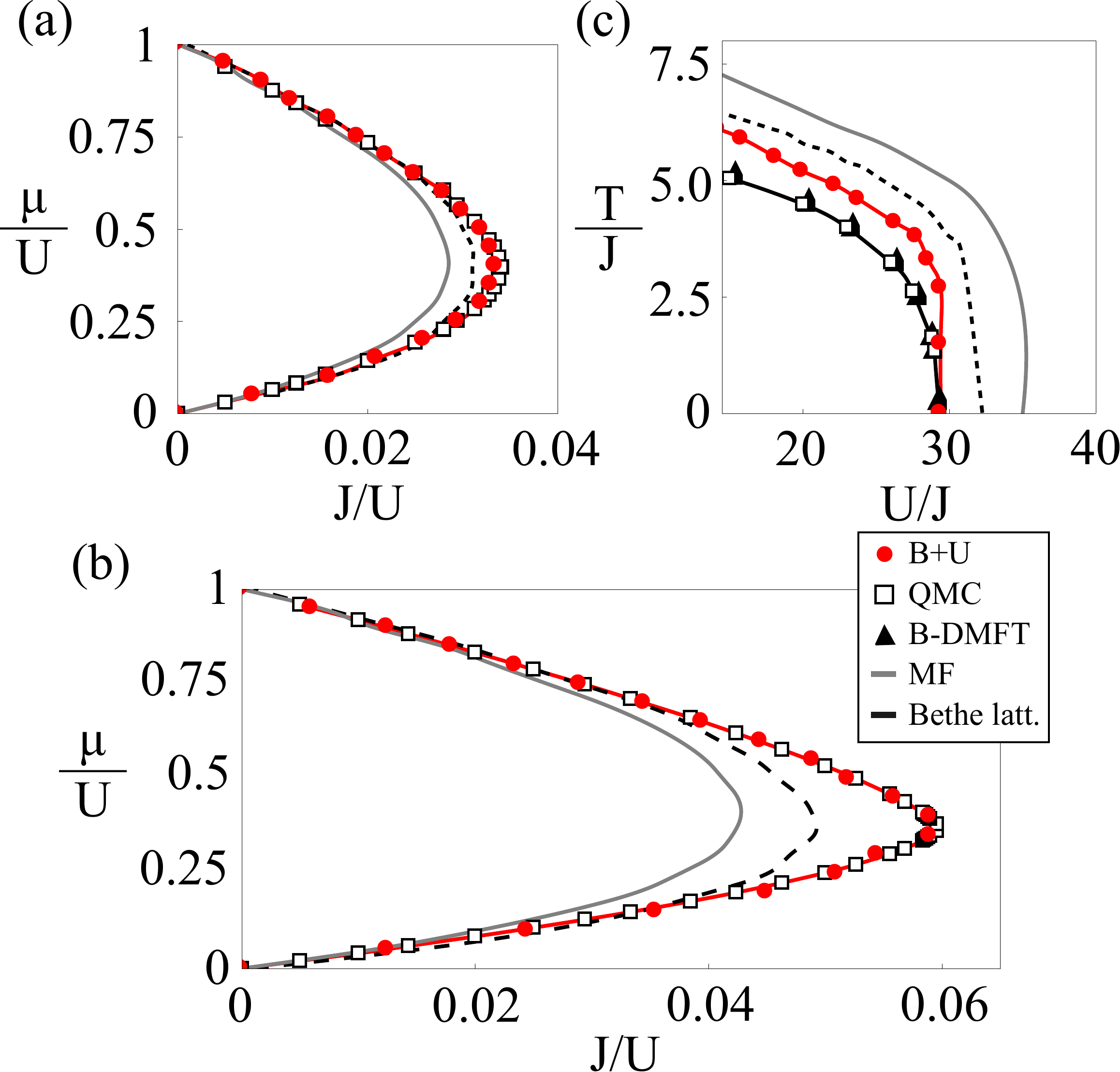}
\caption{\label{fig1} (Color online) {(a),(b) Zero-temperature phase diagram of a (a) 3D cubic and (b) 2D square lattice in the vicinity of the $\langle{n}\rangle=1$ Mott lobe calculated with B+U (red dots) compared with mean-field (gray line) and QMC [\onlinecite{QMC_Cap}] (white boxes) results. For simplicity, the B-DMFT results are not shown here, since they overlap with the QMC data within $1 \%$ [\onlinecite{BDMFT}].  (c) Temperature-dependent phase diagram of a 3D cubic lattice with chemical potential  $\mu/U=0.4$ calculated with B+U (red dots), compared with mean-field (gray line), QMC [\onlinecite{QMC_Cap}] (white boxes, mostly overlapping with B-DMFT) and B-DMFT [\onlinecite{BDMFT}] (black triangles) results. The B+U results for a semi-circular density of states (Bethe lattice, $z=6$) are shown as a black dashed line. The systematic error bar is smaller than the size of the dots.}}
\end{figure}

\begin{figure}
\includegraphics[scale=.35]{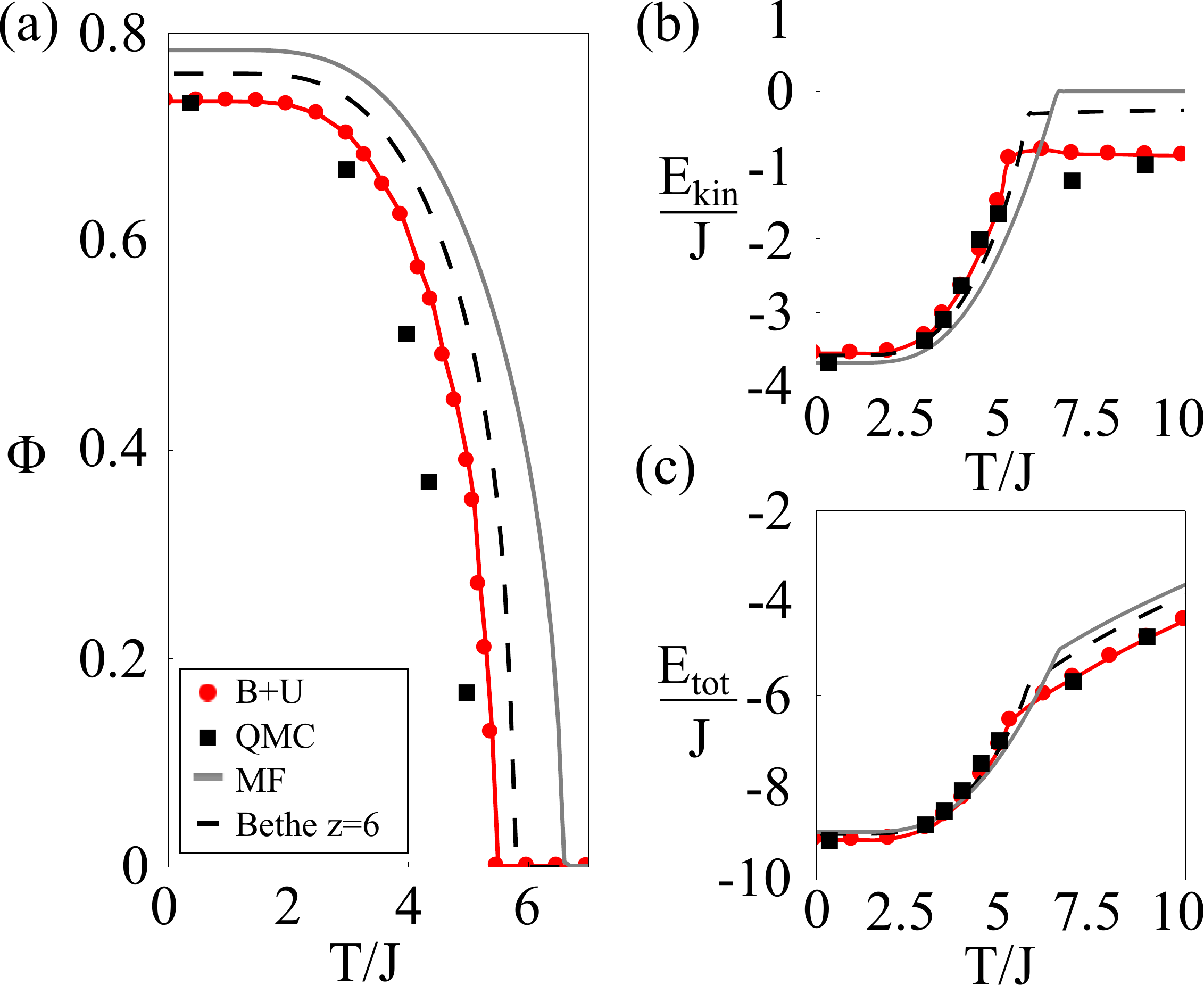}
\caption{\label{fig2} (Color online) {Temperature dependence of (a) the superfluid order parameter $\phi$ , (b) the kinetic energy $E_{\rm kin}$, and (c) the total energy $E_{\rm tot}$ for $\mu/U=0.4$ ($\langle{n}\rangle\approx{1}$) and $U/J=20$ in a 3D cubic lattice calculated with B+U (red dots) compared with mean-field (gray line) and QMC [\onlinecite{QMC_Cap}] (black boxes) results. The B+U results for a semicircular density of states (Bethe lattice, $z=6$) are shown as a black dashed line. The systematic error bar is smaller than the size of the dots.}}
\end{figure}
\begin{figure}
\includegraphics[scale=.25]{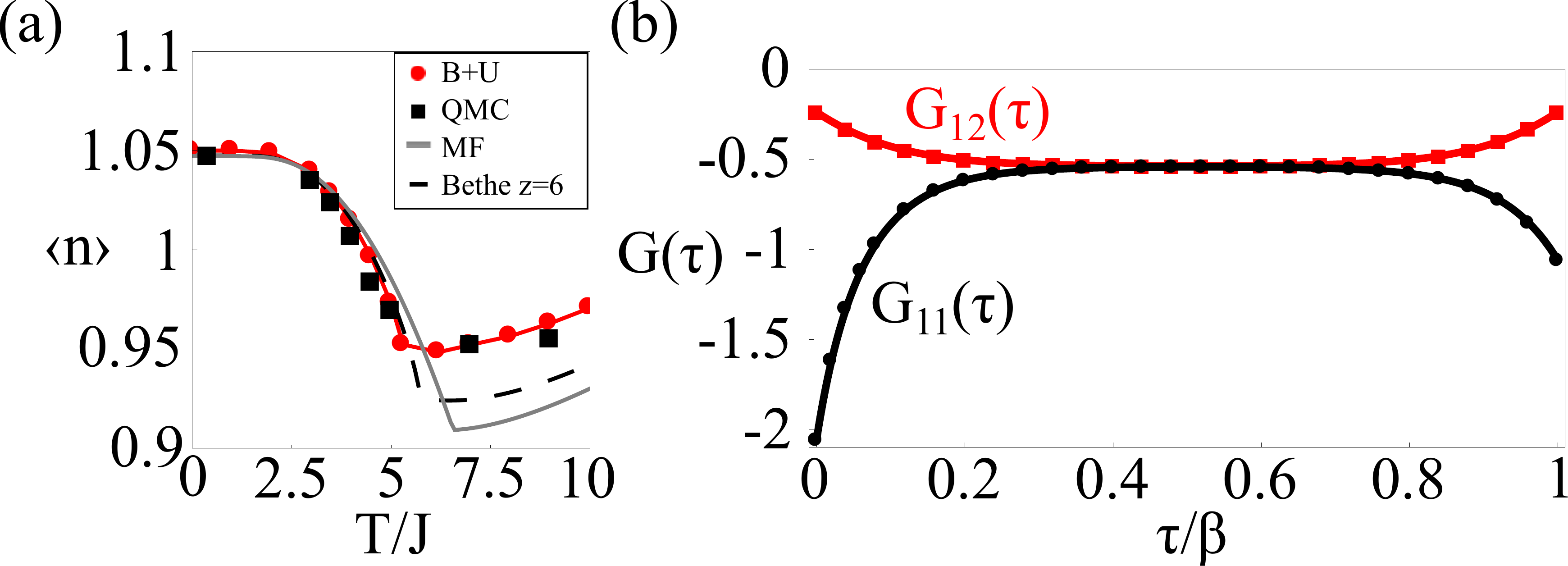}
\caption{\label{fig3} (Color online) {(a) Temperature dependence of the local density per site $\langle n \rangle$ for $\mu/U=0.4$ and $U/J=20$ in a 3D cubic lattice calculated with B+U (red dots) compared with mean-field (gray line), Bethe lattice (dashed black line), and QMC [\onlinecite{QMC_Cap}] (black boxes) results. (b) Imaginary time dependence of the components of the Green's function on the impurity ${\bf G}(\tau)$ in the superfluid phase for the same parameters and $T/J=1$. The normal component $G_{11}(\tau)=-\langle b^{}_{o}(\tau)b^{\dagger}_{o}(0) \rangle$ is shown in black dots , while the anomalous component $G_{12}(\tau)=-\langle b^{}_{o}(\tau)b^{}_{o}(0) \rangle$ is shown in red squares. }}
\end{figure}
\begin{figure}
\includegraphics[scale=.35]{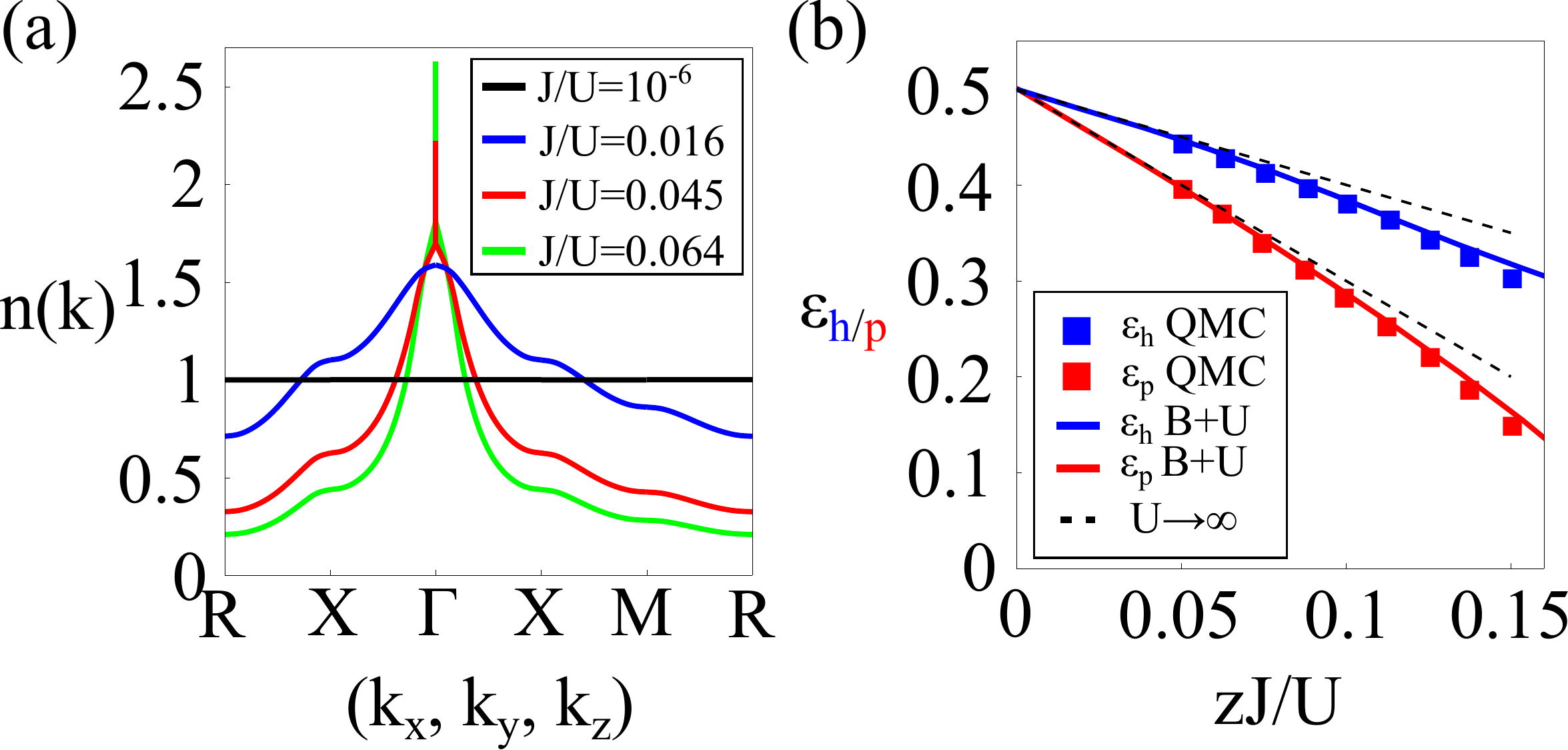}
\caption{\label{fig4} (Color online) {(a) Momentum dependence of the density $n(k)$ at zero temperature for $\mu/U=0.5$ and different values of $J/U$ in the Mott (black, blue) and superfluid phase (red, green). (b) Quasi particle and quasi hole energies at zero momentum $\epsilon_{p/h}$ in the Mott phase at zero temperature for B+U (solid lines), QMC [\onlinecite{Duchon}] (squares), and the analytic zeroth-order solution for $U\rightarrow\infty$ (dashed) for $\mu/U=0.5$ and zero temperature. }}
\end{figure}
In Fig. \ref{fig1} the phase diagram at zero temperature is shown for both a three-dimensional (3D) cubic and a two-dimensional (2D) square lattice. We compare the results with mean-field theory, path integral Monte Carlo simulations with worm-type updates (QMC) from Ref. [\onlinecite{QMC_Cap}], and B-DMFT results from Ref. [\onlinecite{BDMFT}]. The results are identical with the B-DMFT results and agree within a percent with the QMC data both for the 3D and the 2D cases. The results for a Bethe lattice  with coordination number $z=6$ are shown as black dashed lines. As can be seen, for the 3D case the simplified self-consistency for the Bethe lattice works very well, showing deviations only near the tip of the Mott lobe. In Fig. \ref{fig1}(c) the temperature-dependent phase diagram for $\mu/U=0.4$ is shown and compared to B-DMFT, QMC, and mean-field results for a 3D cubic lattice. In this case the lack of retardation in the B+U formalism leads to a bigger deviation from the B-DMFT results. However, the B+U results are still far more precise than the ones obtained in static mean field theory. In Fig. \ref{fig2} the temperature dependence of the superfluid order parameter $\phi$, the kinetic energy $E_{\rm kin}$, and the total energy $E_{\rm tot}$ is shown and compared to mean-field and QMC for $\mu/U=0.4$ ($\langle{n}\rangle\approx{1}$) and $U/J=20$. It should be noted that, since the optimization in ${\bf{\Sigma}}$ is very sensitive close to the phase transition, we cannot and wish not to make any statements with respect to the order of the phase transition in Fig. \ref{fig2}: A  local theory such as B+U should be judged for its accuracy on local observables and is by construction unable to capture long wavelength physics. Information on critical phenomena is hence outside its realm of applicability. The kinetic energy is very accurate for low temperatures, but in the normal phase we find a plateau just as in the decoupling approximation. The corresponding local density $\langle n \rangle$ per site is shown in Fig. \ref{fig3} for the same parameters, where also the full local Green's function on the impurity in imaginary time ${\bf G}(\tau)=-\langle \Phi^{}_o(\tau)\Phi^{\dagger}_o(0) \rangle_{H_{E}}$ calculated by (\ref{formelt}) is shown for the same parameters and temperature $T/J=1$. In Fig. \ref{fig4}(a) we plot the density in momentum-space $n(k)=n_{\rm dp}(k)+n_0 \delta_{k,0}$ for $\mu/U=0.5$ and different values of $J/U$ at zero temperature, where $n_{\rm dp}(k)$ are the depleted particles calculated from the connected Green's function and $n_0 = |\phi|^2$ is the condensate fraction. In Fig. \ref{fig4}(b) the quasi-particle and quasi-hole energies in the Mott phase at zero momentum are extrapolated from the imaginary-time dependence of the zero-momentum Green's function through (\ref{formels4}) and compared to QMC results from Ref. [\onlinecite{Duchon}] and the analytic zeroth-order solution for $U\rightarrow\infty$. We see both good agreement with QMC for finite $J/U$ as with the strongly interacting limit for $J/U\ll 1$

\section{Conclusion}
\label{sec6}

We have presented the B+U framework for equilibrium studies of the Bose-Hubbard model. It captures the low-energy physics of a condensed phase as well as a Mott localization transition when density fluctuations are strongly suppressed on a lattice. The thermodynamics of local quantities of the Bose-Hubbard model can be accurately reproduced everywhere in parameter space by just three parameters $\phi$, $\Sigma_{11}$, and $\Sigma_{12}$. By treating the self-energy as a variational parameter which minimizes the grand potential, B+U can reproduce both the 3D and 2D phase transition from the Mott to the superfluid phase at zero temperature with an accuracy of about $1\%$ near the tip of the lobe and better elsewhere. The B+U method can also be applied to finite-temperature systems showing a strong improvement on the simple mean-field limit but it is less accurate than B-DMFT in locating the phase transition line. Just as B-DMFT,  being a local theory, it can of course never capture hydrodynamics. Due to its simplicity and low computational cost, B+U is a powerful tool which in the future could be extended to clusters in order to study inhomogeneous or topologically non trivial bosonic systems, combining exact interactions on clusters, embedded self-consistently in a lattice with any dispersion.

\begin{acknowledgments}
We wish to thank M. Eckstein, H. U. R. Strand, and F. A. Wolf for fruitful discussions. This work was supported by FP7/ERC Starting Grant No. 306897 and FP7/Marie-Curie Grant No. 321918.
\end{acknowledgments}
\begin{widetext}
\appendix*

\section{Relation between B+U and B-DMFT}
\label{Ap1}
We start from the Hamiltonian (\ref{formela}), but instead of $Z={\rm tr}\left[e^{-\beta H}\right]$ we treat the partition function in the presence of retardation as $Z={\rm tr}\left[e^{-S}\right]$ with the full action $S$ given by

\begin{equation}
 \label{formela1}
S =  \int_0^{\beta} d\tau \left( \sum_i b_i^{\dagger}(\tau) \partial_{\tau} b_i(\tau) - J\sum_{\langle{i,j}\rangle} b^{\dagger}_{i}(\tau) b_{j}(\tau)\right)+ \int_0^{\beta} d\tau\sum_i  \left(\frac{U}{2} n_i(\tau)(n_i (\tau)- 1) - \mu n_i(\tau) \right).
\end{equation}\\

As in Sec. \ref{sec2} we expand $b_{j}(\tau)=\phi+\delta b_{j}(\tau)$ around its site- and imaginary-time-independent mean-field value $\phi$, giving us

\begin{eqnarray}
S & = &  S_o + S_{\rm ext} + \Delta S, \nonumber \\
S_o & = & \int_0^{\beta} d\tau \left( b_o^{\dagger}(\tau) \partial_{\tau} b_o(\tau) + \frac{U}{2}  n_o(\tau)(n_o (\tau)- 1) - \mu n_o(\tau) -zJ \phi (b^{\dagger}_o(\tau)+b_o(\tau))\right), \label{formela2}  \\ 
\Delta S & = & - J\int_0^{\beta} d\tau\sum_{\langle{i,o}\rangle}\left(\delta b^{\dagger}_{i}(\tau)\delta b_{o}(\tau)+\delta b^{\dagger}_{o}(\tau)\delta b_{i}(\tau)\right),   \nonumber
\end{eqnarray}

and expand the full partition function as  $Z = Z_{\rm ext} Z_o$ by
\begin{equation}
\label{formela3} 
Z_o =  {\rm tr}\left[e^{-S_o-\left< \Delta S\right>_{S_{\rm ext}} }  \right]
\end{equation}
As with the Hamiltonian in Sec. \ref{sec2} we approximate the expectation value $\langle\Delta S\rangle_{S_{\rm ext}}$ by the cumulant expansion,

\begin{eqnarray}
\left< \Delta S \right>_{S_{\rm ext}} & \approx & -\int_0^{\beta} d\tau J\left\langle\sum_{\langle{i,o}\rangle}\delta \Phi^{\dagger}_{i}(\tau)\delta \Phi_{o}(\tau)\right\rangle_{S_{\rm ext}}-\frac{1}{2}\int_0^{\beta} d\tau d\tau'J^{2}\left\langle\sum_{\langle{i,o}\rangle}\delta\Phi^{\dagger}_{i}(\tau)\delta\Phi_{o}(\tau)\sum_{\langle{j,o}\rangle}\delta\Phi^{\dagger}_{j}(\tau')\delta\Phi_{o}(\tau')\right\rangle_{S_{\rm ext}}  \nonumber \\
 & = & 0-\int_0^{\beta} d\tau d\tau'\frac{1}{2}\delta \Phi^{\dagger}_{o}(\tau){\bf{\Delta}}(\tau-\tau')\delta\Phi_{o}(\tau'), \label{formela4}
\end{eqnarray}

leading to the effective B-DMFT impurity action [\onlinecite{BDMFT},\onlinecite{BDMFT2}]
\begin{eqnarray}
S_E & = & \int_0^{\beta} d\tau\left(b^{\dagger}_o(\tau) \partial_{\tau} b_o(\tau) -zJ\phi(b_o(\tau)+b_o^{\dagger}(\tau))\right)\ -\int_0^{\beta} d\tau d\tau'\frac{1}{2}\delta \Phi^{\dagger}_{o}(\tau){\bf{\Delta}}(\tau-\tau')\delta\Phi_{o}(\tau') \nonumber  \\
{} & {} &  + \int_0^{\beta} d\tau \left(\frac{U}{2}  n_o(\tau) (n_o(\tau) - 1) - \mu n_o(\tau)\right).  \label{formela5}
\end{eqnarray}
By the relation 
$\\ S[b,b^{\dagger}]= \int_0^{\beta} d\tau\left(b^{\dagger}(\tau) \partial_{\tau} b(\tau)+H[b(\tau),b^{\dagger}(\tau)]\right)$, one can see that the action (\ref{formela5}) is equivalent to the effective Hamiltonian (\ref{formelf}) for the ansatz
\begin{equation}
{\bf \Delta}(\tau - \tau') \to {\bf \Delta} \delta(\tau - \tau')
\label{formela6}
\end{equation}
In fact, (\ref{formela6}) reduces the Dyson equation of B-DMFT on the impurity to
\begin{equation}
i\omega_n  \sigma_z+\mu{\bf I}+ {\bf \Delta}-{\bf \Sigma}(i\omega_n)={\bf G_c}^{-1}(i\omega_n).
\label{formela8}
\end{equation}
If we now take the zero Matsubara frequency of this equation we recover the B+U Dyson equation on the impurity  (\ref{formelh}).

\end{widetext}

\bibliography{Paper_PRB2}

\begin{thebibliography}{29}%
\makeatletter
\providecommand \@ifxundefined [1]{%
 \@ifx{#1\undefined}
}%
\providecommand \@ifnum [1]{%
 \ifnum #1\expandafter \@firstoftwo
 \else \expandafter \@secondoftwo
 \fi
}%
\providecommand \@ifx [1]{%
 \ifx #1\expandafter \@firstoftwo
 \else \expandafter \@secondoftwo
 \fi
}%
\providecommand \natexlab [1]{#1}%
\providecommand \enquote  [1]{``#1''}%
\providecommand \bibnamefont  [1]{#1}%
\providecommand \bibfnamefont [1]{#1}%
\providecommand \citenamefont [1]{#1}%
\providecommand \href@noop [0]{\@secondoftwo}%
\providecommand \href [0]{\begingroup \@sanitize@url \@href}%
\providecommand \@href[1]{\@@startlink{#1}\@@href}%
\providecommand \@@href[1]{\endgroup#1\@@endlink}%
\providecommand \@sanitize@url [0]{\catcode `\\12\catcode `\$12\catcode
  `\&12\catcode `\#12\catcode `\^12\catcode `\_12\catcode `\%12\relax}%
\providecommand \@@startlink[1]{}%
\providecommand \@@endlink[0]{}%
\providecommand \url  [0]{\begingroup\@sanitize@url \@url }%
\providecommand \@url [1]{\endgroup\@href {#1}{\urlprefix }}%
\providecommand \urlprefix  [0]{URL }%
\providecommand \Eprint [0]{\href }%
\providecommand \doibase [0]{http://dx.doi.org/}%
\providecommand \selectlanguage [0]{\@gobble}%
\providecommand \bibinfo  [0]{\@secondoftwo}%
\providecommand \bibfield  [0]{\@secondoftwo}%
\providecommand \translation [1]{[#1]}%
\providecommand \BibitemOpen [0]{}%
\providecommand \bibitemStop [0]{}%
\providecommand \bibitemNoStop [0]{.\EOS\space}%
\providecommand \EOS [0]{\spacefactor3000\relax}%
\providecommand \BibitemShut  [1]{\csname bibitem#1\endcsname}%
\let\auto@bib@innerbib\@empty
\bibitem [{\citenamefont {Bloch}\ \emph {et~al.}(2008)\citenamefont {Bloch},
  \citenamefont {Dalibard},\ and\ \citenamefont {Zwerger}}]{BlochRev}%
  \BibitemOpen
  \bibfield  {author} {\bibinfo {author} {\bibfnamefont {I.}~\bibnamefont
  {Bloch}}, \bibinfo {author} {\bibfnamefont {J.}~\bibnamefont {Dalibard}}, \
  and\ \bibinfo {author} {\bibfnamefont {W.}~\bibnamefont {Zwerger}},\
  }\href@noop {} {\bibfield  {journal} {\bibinfo  {journal} {Rev. of Mod.
  Phys.}\ }\textbf {\bibinfo {volume} {80}},\ \bibinfo {pages} {885} (\bibinfo
  {year} {2008})}\BibitemShut {NoStop}%
\bibitem [{\citenamefont {Aidelsburger}\ \emph {et~al.}(2013)\citenamefont
  {Aidelsburger}, \citenamefont {Atala}, \citenamefont {Lohse}, \citenamefont
  {Barreiro}, \citenamefont {Paredes},\ and\ \citenamefont {Bloch}}]{Hof}%
  \BibitemOpen
  \bibfield  {author} {\bibinfo {author} {\bibfnamefont {M.}~\bibnamefont
  {Aidelsburger}}, \bibinfo {author} {\bibfnamefont {M.}~\bibnamefont {Atala}},
  \bibinfo {author} {\bibfnamefont {M.}~\bibnamefont {Lohse}}, \bibinfo
  {author} {\bibfnamefont {J.~T.}\ \bibnamefont {Barreiro}}, \bibinfo {author}
  {\bibfnamefont {B.}~\bibnamefont {Paredes}}, \ and\ \bibinfo {author}
  {\bibfnamefont {I.}~\bibnamefont {Bloch}},\ }\href@noop {} {\bibfield
  {journal} {\bibinfo  {journal} {Phys. Rev. Lett.}\ }\textbf {\bibinfo
  {volume} {111}},\ \bibinfo {pages} {185301} (\bibinfo {year}
  {2013})}\BibitemShut {NoStop}%
\bibitem [{\citenamefont {K\"{u}hner}\ and\ \citenamefont
  {Monien}(1998)}]{Kuhner}%
  \BibitemOpen
  \bibfield  {author} {\bibinfo {author} {\bibfnamefont {T.~D.}\ \bibnamefont
  {K\"{u}hner}}\ and\ \bibinfo {author} {\bibfnamefont {H.}~\bibnamefont
  {Monien}},\ }\href@noop {} {\bibfield  {journal} {\bibinfo  {journal} {Phys.
  Rev. B}\ }\textbf {\bibinfo {volume} {58}},\ \bibinfo {pages} {R14741}
  (\bibinfo {year} {1998})}\BibitemShut {NoStop}%
\bibitem [{\citenamefont {Rapsch}\ \emph {et~al.}(1999)\citenamefont {Rapsch},
  \citenamefont {Schollw\"{o}ck},\ and\ \citenamefont {Zwerger}}]{Rapsch}%
  \BibitemOpen
  \bibfield  {author} {\bibinfo {author} {\bibfnamefont {S.}~\bibnamefont
  {Rapsch}}, \bibinfo {author} {\bibfnamefont {U.}~\bibnamefont
  {Schollw\"{o}ck}}, \ and\ \bibinfo {author} {\bibfnamefont {W.}~\bibnamefont
  {Zwerger}},\ }\href@noop {} {\bibfield  {journal} {\bibinfo  {journal}
  {Europhys. Lett.}\ }\textbf {\bibinfo {volume} {46}},\ \bibinfo {pages} {559}
  (\bibinfo {year} {1999})}\BibitemShut {NoStop}%
\bibitem [{\citenamefont {Kollath}\ \emph {et~al.}(2004)\citenamefont
  {Kollath}, \citenamefont {Schollw\"{o}ck}, \citenamefont {von Delft},\ and\
  \citenamefont {Zwerger}}]{Kollath1}%
  \BibitemOpen
  \bibfield  {author} {\bibinfo {author} {\bibfnamefont {C.}~\bibnamefont
  {Kollath}}, \bibinfo {author} {\bibfnamefont {U.}~\bibnamefont
  {Schollw\"{o}ck}}, \bibinfo {author} {\bibfnamefont {J.}~\bibnamefont {von
  Delft}}, \ and\ \bibinfo {author} {\bibfnamefont {W.}~\bibnamefont
  {Zwerger}},\ }\href@noop {} {\bibfield  {journal} {\bibinfo  {journal} {Phys.
  Rev. A}\ }\textbf {\bibinfo {volume} {69}},\ \bibinfo {pages} {031601(R)}
  (\bibinfo {year} {2004})}\BibitemShut {NoStop}%
\bibitem [{\citenamefont {Kollath}\ \emph {et~al.}(2007)\citenamefont
  {Kollath}, \citenamefont {L\"{a}uchli},\ and\ \citenamefont
  {Altman}}]{Kollath2}%
  \BibitemOpen
  \bibfield  {author} {\bibinfo {author} {\bibfnamefont {C.}~\bibnamefont
  {Kollath}}, \bibinfo {author} {\bibfnamefont {A.~M.}\ \bibnamefont
  {L\"{a}uchli}}, \ and\ \bibinfo {author} {\bibfnamefont {E.}~\bibnamefont
  {Altman}},\ }\href@noop {} {\bibfield  {journal} {\bibinfo  {journal} {Phys.
  Rev. Lett.}\ }\textbf {\bibinfo {volume} {98}},\ \bibinfo {pages} {180601}
  (\bibinfo {year} {2007})}\BibitemShut {NoStop}%
\bibitem [{\citenamefont {Prokof'ev}\ \emph {et~al.}(1998)\citenamefont
  {Prokof'ev}, \citenamefont {Svistunov},\ and\ \citenamefont
  {Tupitsyn}}]{Tupitsyn}%
  \BibitemOpen
  \bibfield  {author} {\bibinfo {author} {\bibfnamefont {N.}~\bibnamefont
  {Prokof'ev}}, \bibinfo {author} {\bibfnamefont {B.}~\bibnamefont
  {Svistunov}}, \ and\ \bibinfo {author} {\bibfnamefont {I.}~\bibnamefont
  {Tupitsyn}},\ }\href@noop {} {\bibfield  {journal} {\bibinfo  {journal} {J.
  Exp. Theor. Phys.}\ }\textbf {\bibinfo {volume} {87}},\ \bibinfo {pages}
  {310} (\bibinfo {year} {1998})}\BibitemShut {NoStop}%
\bibitem [{\citenamefont {Trotzky}\ \emph {et~al.}(2010)\citenamefont
  {Trotzky}, \citenamefont {Pollet}, \citenamefont {Gerbier}, \citenamefont
  {Schnorrberger}, \citenamefont {Bloch}, \citenamefont {Prokof'ev},
  \citenamefont {Svistunov},\ and\ \citenamefont {Troyer}}]{Trotzky}%
  \BibitemOpen
  \bibfield  {author} {\bibinfo {author} {\bibfnamefont {S.}~\bibnamefont
  {Trotzky}}, \bibinfo {author} {\bibfnamefont {L.}~\bibnamefont {Pollet}},
  \bibinfo {author} {\bibfnamefont {F.}~\bibnamefont {Gerbier}}, \bibinfo
  {author} {\bibfnamefont {U.}~\bibnamefont {Schnorrberger}}, \bibinfo {author}
  {\bibfnamefont {I.}~\bibnamefont {Bloch}}, \bibinfo {author} {\bibfnamefont
  {N.}~\bibnamefont {Prokof'ev}}, \bibinfo {author} {\bibfnamefont
  {B.}~\bibnamefont {Svistunov}}, \ and\ \bibinfo {author} {\bibfnamefont
  {M.}~\bibnamefont {Troyer}},\ }\href@noop {} {\bibfield  {journal} {\bibinfo
  {journal} {Nat. Phys.}\ }\textbf {\bibinfo {volume} {6}},\ \bibinfo {pages}
  {998} (\bibinfo {year} {2010})}\BibitemShut {NoStop}%
\bibitem [{\citenamefont {Pollet}(2012)}]{Rev_Lode}%
  \BibitemOpen
  \bibfield  {author} {\bibinfo {author} {\bibfnamefont {L.}~\bibnamefont
  {Pollet}},\ }\href@noop {} {\bibfield  {journal} {\bibinfo  {journal} {Rep.
  Prog. Phys.}\ }\textbf {\bibinfo {volume} {75}},\ \bibinfo {pages} {094501}
  (\bibinfo {year} {2012})}\BibitemShut {NoStop}%
\bibitem [{\citenamefont {Anders}\ \emph {et~al.}(2011)\citenamefont {Anders},
  \citenamefont {Gull}, \citenamefont {Pollet}, \citenamefont {Troyer},\ and\
  \citenamefont {Werner}}]{BDMFT}%
  \BibitemOpen
  \bibfield  {author} {\bibinfo {author} {\bibfnamefont {P.}~\bibnamefont
  {Anders}}, \bibinfo {author} {\bibfnamefont {E.}~\bibnamefont {Gull}},
  \bibinfo {author} {\bibfnamefont {L.}~\bibnamefont {Pollet}}, \bibinfo
  {author} {\bibfnamefont {M.}~\bibnamefont {Troyer}}, \ and\ \bibinfo {author}
  {\bibfnamefont {P.}~\bibnamefont {Werner}},\ }\href@noop {} {\bibfield
  {journal} {\bibinfo  {journal} {New J. Phys.}\ }\textbf {\bibinfo {volume}
  {13}},\ \bibinfo {pages} {075013} (\bibinfo {year} {2011})}\BibitemShut
  {NoStop}%
\bibitem [{\citenamefont {Anders}\ \emph {et~al.}(2010)\citenamefont {Anders},
  \citenamefont {Gull}, \citenamefont {Pollet}, \citenamefont {Troyer},\ and\
  \citenamefont {Werner}}]{BDMFT2}%
  \BibitemOpen
  \bibfield  {author} {\bibinfo {author} {\bibfnamefont {P.}~\bibnamefont
  {Anders}}, \bibinfo {author} {\bibfnamefont {E.}~\bibnamefont {Gull}},
  \bibinfo {author} {\bibfnamefont {L.}~\bibnamefont {Pollet}}, \bibinfo
  {author} {\bibfnamefont {M.}~\bibnamefont {Troyer}}, \ and\ \bibinfo {author}
  {\bibfnamefont {P.}~\bibnamefont {Werner}},\ }\href@noop {} {\bibfield
  {journal} {\bibinfo  {journal} {Phys. Rev. Lett.}\ }\textbf {\bibinfo
  {volume} {105}},\ \bibinfo {pages} {096402} (\bibinfo {year}
  {2010})}\BibitemShut {NoStop}%
\bibitem [{\citenamefont {He}\ \emph {et~al.}(2014)\citenamefont {He},
  \citenamefont {Ji},\ and\ \citenamefont {Hofstetter}}]{Hofst1}%
  \BibitemOpen
  \bibfield  {author} {\bibinfo {author} {\bibfnamefont {L.}~\bibnamefont
  {He}}, \bibinfo {author} {\bibfnamefont {A.}~\bibnamefont {Ji}}, \ and\
  \bibinfo {author} {\bibfnamefont {W.}~\bibnamefont {Hofstetter}},\
  }\href@noop {} {} (\bibinfo {year} {2014})\BibitemShut {NoStop}%
\bibitem [{\citenamefont {He}\ \emph {et~al.}(2012)\citenamefont {He},
  \citenamefont {Li}, \citenamefont {Altman},\ and\ \citenamefont
  {Hofstetter}}]{Hofst3}%
  \BibitemOpen
  \bibfield  {author} {\bibinfo {author} {\bibfnamefont {L.}~\bibnamefont
  {He}}, \bibinfo {author} {\bibfnamefont {Y.}~\bibnamefont {Li}}, \bibinfo
  {author} {\bibfnamefont {E.}~\bibnamefont {Altman}}, \ and\ \bibinfo {author}
  {\bibfnamefont {W.}~\bibnamefont {Hofstetter}},\ }\href@noop {} {\bibfield
  {journal} {\bibinfo  {journal} {Phys. Rev. A}\ }\textbf {\bibinfo {volume}
  {86}},\ \bibinfo {pages} {043620} (\bibinfo {year} {2012})}\BibitemShut
  {NoStop}%
\bibitem [{\citenamefont {Akerlund}\ and\ \citenamefont
  {de~Forcrand}(2013)}]{EMFT1}%
  \BibitemOpen
  \bibfield  {author} {\bibinfo {author} {\bibfnamefont {O.}~\bibnamefont
  {Akerlund}}\ and\ \bibinfo {author} {\bibfnamefont {P.}~\bibnamefont
  {de~Forcrand}},\ }\href@noop {} {} (\bibinfo {year} {2013})\BibitemShut
  {NoStop}%
\bibitem [{\citenamefont {Akerlund}\ \emph {et~al.}(2014)\citenamefont
  {Akerlund}, \citenamefont {de~Forcrand}, \citenamefont {Georges},\ and\
  \citenamefont {Werner}}]{EMFT2}%
  \BibitemOpen
  \bibfield  {author} {\bibinfo {author} {\bibfnamefont {O.}~\bibnamefont
  {Akerlund}}, \bibinfo {author} {\bibfnamefont {P.}~\bibnamefont
  {de~Forcrand}}, \bibinfo {author} {\bibfnamefont {A.}~\bibnamefont
  {Georges}}, \ and\ \bibinfo {author} {\bibfnamefont {P.}~\bibnamefont
  {Werner}},\ }\href@noop {} {\bibfield  {journal} {\bibinfo  {journal} {Phys.
  Rev. D}\ }\textbf {\bibinfo {volume} {90}},\ \bibinfo {pages} {065008}
  (\bibinfo {year} {2014})}\BibitemShut {NoStop}%
\bibitem [{\citenamefont {Knap}\ \emph {et~al.}(2011)\citenamefont {Knap},
  \citenamefont {Arrigoni},\ and\ \citenamefont {von~der Linden}}]{VCA}%
  \BibitemOpen
  \bibfield  {author} {\bibinfo {author} {\bibfnamefont {M.}~\bibnamefont
  {Knap}}, \bibinfo {author} {\bibfnamefont {E.}~\bibnamefont {Arrigoni}}, \
  and\ \bibinfo {author} {\bibfnamefont {W.}~\bibnamefont {von~der Linden}},\
  }\href@noop {} {\bibfield  {journal} {\bibinfo  {journal} {Phys. Rev. B}\
  }\textbf {\bibinfo {volume} {83}},\ \bibinfo {pages} {134507} (\bibinfo
  {year} {2011})}\BibitemShut {NoStop}%
\bibitem [{\citenamefont {Capogrosso-Sansone}\ \emph
  {et~al.}(2010)\citenamefont {Capogrosso-Sansone}, \citenamefont {Giorgini},
  \citenamefont {Pilati}, \citenamefont {Pollet}, \citenamefont {Prokof'ev},
  \citenamefont {Svistunov},\ and\ \citenamefont {Troyer}}]{Svist-Gior-Poll}%
  \BibitemOpen
  \bibfield  {author} {\bibinfo {author} {\bibfnamefont {B.}~\bibnamefont
  {Capogrosso-Sansone}}, \bibinfo {author} {\bibfnamefont {S.}~\bibnamefont
  {Giorgini}}, \bibinfo {author} {\bibfnamefont {S.}~\bibnamefont {Pilati}},
  \bibinfo {author} {\bibfnamefont {L.}~\bibnamefont {Pollet}}, \bibinfo
  {author} {\bibfnamefont {N.}~\bibnamefont {Prokof'ev}}, \bibinfo {author}
  {\bibfnamefont {B.}~\bibnamefont {Svistunov}}, \ and\ \bibinfo {author}
  {\bibfnamefont {M.}~\bibnamefont {Troyer}},\ }\href@noop {} {\bibfield
  {journal} {\bibinfo  {journal} {New J. Phys.}\ }\textbf {\bibinfo {volume}
  {12}},\ \bibinfo {pages} {043010} (\bibinfo {year} {2010})}\BibitemShut
  {NoStop}%
\bibitem [{\citenamefont {Kleinert}\ \emph {et~al.}(2014)\citenamefont
  {Kleinert}, \citenamefont {Narzikulov},\ and\ \citenamefont
  {Rakhimov}}]{Kleinert}%
  \BibitemOpen
  \bibfield  {author} {\bibinfo {author} {\bibfnamefont {H.}~\bibnamefont
  {Kleinert}}, \bibinfo {author} {\bibfnamefont {Z.}~\bibnamefont
  {Narzikulov}}, \ and\ \bibinfo {author} {\bibfnamefont {A.}~\bibnamefont
  {Rakhimov}},\ }\href@noop {} {\bibfield  {journal} {\bibinfo  {journal} {J.
  Stat. Mech.}\ ,\ \bibinfo {pages} {P01003}} (\bibinfo {year}
  {2014})}\BibitemShut {NoStop}%
\bibitem [{\citenamefont {Dawson}\ \emph {et~al.}(2013)\citenamefont {Dawson},
  \citenamefont {Cooper}, \citenamefont {Chien},\ and\ \citenamefont
  {Mihaila}}]{Cooper1}%
  \BibitemOpen
  \bibfield  {author} {\bibinfo {author} {\bibfnamefont {J.~F.}\ \bibnamefont
  {Dawson}}, \bibinfo {author} {\bibfnamefont {F.}~\bibnamefont {Cooper}},
  \bibinfo {author} {\bibfnamefont {C.-C.}\ \bibnamefont {Chien}}, \ and\
  \bibinfo {author} {\bibfnamefont {B.}~\bibnamefont {Mihaila}},\ }\href@noop
  {} {\bibfield  {journal} {\bibinfo  {journal} {Phys. Rev. A}\ }\textbf
  {\bibinfo {volume} {88}},\ \bibinfo {pages} {023607} (\bibinfo {year}
  {2013})}\BibitemShut {NoStop}%
\bibitem [{\citenamefont {Cooper}\ \emph {et~al.}(2012)\citenamefont {Cooper},
  \citenamefont {Chien}, \citenamefont {Mihaila}, \citenamefont {Dawson},\ and\
  \citenamefont {Timmermans}}]{Cooper2}%
  \BibitemOpen
  \bibfield  {author} {\bibinfo {author} {\bibfnamefont {F.}~\bibnamefont
  {Cooper}}, \bibinfo {author} {\bibfnamefont {C.-C.}\ \bibnamefont {Chien}},
  \bibinfo {author} {\bibfnamefont {B.}~\bibnamefont {Mihaila}}, \bibinfo
  {author} {\bibfnamefont {J.~F.}\ \bibnamefont {Dawson}}, \ and\ \bibinfo
  {author} {\bibfnamefont {E.}~\bibnamefont {Timmermans}},\ }\href@noop {}
  {\bibfield  {journal} {\bibinfo  {journal} {Phys. Rev. A}\ }\textbf {\bibinfo
  {volume} {85}},\ \bibinfo {pages} {023631} (\bibinfo {year}
  {2012})}\BibitemShut {NoStop}%
\bibitem [{\citenamefont {Mihaila}\ \emph {et~al.}(2011)\citenamefont
  {Mihaila}, \citenamefont {Cooper}, \citenamefont {Dawson}, \citenamefont
  {Chien},\ and\ \citenamefont {Timmermans}}]{Cooper3}%
  \BibitemOpen
  \bibfield  {author} {\bibinfo {author} {\bibfnamefont {B.}~\bibnamefont
  {Mihaila}}, \bibinfo {author} {\bibfnamefont {F.}~\bibnamefont {Cooper}},
  \bibinfo {author} {\bibfnamefont {J.}~\bibnamefont {Dawson}}, \bibinfo
  {author} {\bibfnamefont {C.-C.}\ \bibnamefont {Chien}}, \ and\ \bibinfo
  {author} {\bibfnamefont {E.}~\bibnamefont {Timmermans}},\ }\href@noop {}
  {\bibfield  {journal} {\bibinfo  {journal} {Phys. Rev. A}\ }\textbf {\bibinfo
  {volume} {84}},\ \bibinfo {pages} {023603} (\bibinfo {year}
  {2011})}\BibitemShut {NoStop}%
\bibitem [{\citenamefont {Trefzger}\ and\ \citenamefont
  {Sengupta}(2011)}]{Trefzger}%
  \BibitemOpen
  \bibfield  {author} {\bibinfo {author} {\bibfnamefont {C.}~\bibnamefont
  {Trefzger}}\ and\ \bibinfo {author} {\bibfnamefont {K.}~\bibnamefont
  {Sengupta}},\ }\href@noop {} {\bibfield  {journal} {\bibinfo  {journal}
  {Phys. Rev. Lett.}\ }\textbf {\bibinfo {volume} {106}},\ \bibinfo {pages}
  {095702} (\bibinfo {year} {2011})}\BibitemShut {NoStop}%
\bibitem [{\citenamefont {Eckardt}(2009)}]{Eckardt}%
  \BibitemOpen
  \bibfield  {author} {\bibinfo {author} {\bibfnamefont {A.}~\bibnamefont
  {Eckardt}},\ }\href@noop {} {\bibfield  {journal} {\bibinfo  {journal} {Phys.
  Rev. B}\ }\textbf {\bibinfo {volume} {79}},\ \bibinfo {pages} {195131}
  (\bibinfo {year} {2009})}\BibitemShut {NoStop}%
\bibitem [{\citenamefont {Teichmann}\ \emph {et~al.}(2009)\citenamefont
  {Teichmann}, \citenamefont {Hinrichs}, \citenamefont {Holthaus},\ and\
  \citenamefont {Eckardt}}]{Teichmann}%
  \BibitemOpen
  \bibfield  {author} {\bibinfo {author} {\bibfnamefont {N.}~\bibnamefont
  {Teichmann}}, \bibinfo {author} {\bibfnamefont {D.}~\bibnamefont {Hinrichs}},
  \bibinfo {author} {\bibfnamefont {M.}~\bibnamefont {Holthaus}}, \ and\
  \bibinfo {author} {\bibfnamefont {A.}~\bibnamefont {Eckardt}},\ }\href@noop
  {} {\bibfield  {journal} {\bibinfo  {journal} {Phys. Rev. B}\ }\textbf
  {\bibinfo {volume} {79}},\ \bibinfo {pages} {224515} (\bibinfo {year}
  {2009})}\BibitemShut {NoStop}%
\bibitem [{\citenamefont {Polak}\ and\ \citenamefont
  {Kope\'{c}}(2009)}]{Polak}%
  \BibitemOpen
  \bibfield  {author} {\bibinfo {author} {\bibfnamefont {T.~P.}\ \bibnamefont
  {Polak}}\ and\ \bibinfo {author} {\bibfnamefont {T.~K.}\ \bibnamefont
  {Kope\'{c}}},\ }\href@noop {} {\bibfield  {journal} {\bibinfo  {journal} {J.
  Phys. B}\ }\textbf {\bibinfo {volume} {42}},\ \bibinfo {pages} {095302}
  (\bibinfo {year} {2009})}\BibitemShut {NoStop}%
\bibitem [{\citenamefont {Kozik}\ \emph {et~al.}(2015)\citenamefont {Kozik},
  \citenamefont {Ferrero},\ and\ \citenamefont {Georges}}]{Kozik}%
  \BibitemOpen
  \bibfield  {author} {\bibinfo {author} {\bibfnamefont {E.}~\bibnamefont
  {Kozik}}, \bibinfo {author} {\bibfnamefont {M.}~\bibnamefont {Ferrero}}, \
  and\ \bibinfo {author} {\bibfnamefont {A.}~\bibnamefont {Georges}},\
  }\href@noop {} {\bibfield  {journal} {\bibinfo  {journal} {Phys. Rev. Lett.}\
  }\textbf {\bibinfo {volume} {114}},\ \bibinfo {pages} {156402} (\bibinfo
  {year} {2015})}\BibitemShut {NoStop}%
\bibitem [{\citenamefont {Potthoff}(2012)}]{SFT}%
  \BibitemOpen
  \bibfield  {author} {\bibinfo {author} {\bibfnamefont {M.}~\bibnamefont
  {Potthoff}},\ }in\ \href@noop {} {\emph {\bibinfo {booktitle} {Strongly
  Correlated Systems}}},\ \bibinfo {series} {Springer Series Solid State
  Physics}, Vol.\ \bibinfo {volume} {171},\ \bibinfo {editor} {edited by\
  \bibinfo {editor} {\bibfnamefont {A.}~\bibnamefont {Avella}}\ and\ \bibinfo
  {editor} {\bibfnamefont {F.}~\bibnamefont {Mancini}}}\ (\bibinfo  {publisher}
  {Springer},\ \bibinfo {address} {Berlin, Heidelberg},\ \bibinfo {year}
  {2012})\ Chap.~\bibinfo {chapter} {10}, pp.\ \bibinfo {pages}
  {303--339}\BibitemShut {NoStop}%
\bibitem [{\citenamefont {Capogrosso-Sansone}\ \emph
  {et~al.}(2007)\citenamefont {Capogrosso-Sansone}, \citenamefont {Prokof'ev},\
  and\ \citenamefont {Svistunov}}]{QMC_Cap}%
  \BibitemOpen
  \bibfield  {author} {\bibinfo {author} {\bibfnamefont {B.}~\bibnamefont
  {Capogrosso-Sansone}}, \bibinfo {author} {\bibfnamefont {N.}~\bibnamefont
  {Prokof'ev}}, \ and\ \bibinfo {author} {\bibfnamefont {B.}~\bibnamefont
  {Svistunov}},\ }\href@noop {} {\bibfield  {journal} {\bibinfo  {journal}
  {Phys. Rev. B}\ }\textbf {\bibinfo {volume} {75}},\ \bibinfo {pages} {134302}
  (\bibinfo {year} {2007})}\BibitemShut {NoStop}%
\bibitem [{\citenamefont {Duchon}\ \emph {et~al.}(2013)\citenamefont {Duchon},
  \citenamefont {Loh},\ and\ \citenamefont {Trivedi}}]{Duchon}%
  \BibitemOpen
  \bibfield  {author} {\bibinfo {author} {\bibfnamefont {E.}~\bibnamefont
  {Duchon}}, \bibinfo {author} {\bibfnamefont {Y.~L.}\ \bibnamefont {Loh}}, \
  and\ \bibinfo {author} {\bibfnamefont {N.}~\bibnamefont {Trivedi}},\ }in\
  \href@noop {} {\emph {\bibinfo {booktitle} {Novel Superfluids}}},\
  Vol.~\bibinfo {volume} {2},\ \bibinfo {editor} {edited by\ \bibinfo {editor}
  {\bibfnamefont {J.}~\bibnamefont {Ketterson}}\ and\ \bibinfo {editor}
  {\bibfnamefont {K.-H.}\ \bibnamefont {Benneman}}}\ (\bibinfo  {publisher}
  {Oxford University Press},\ \bibinfo {address} {Oxford, UK},\ \bibinfo {year}
  {2013})\ p.\ \bibinfo {pages} {193}\BibitemShut {NoStop}%
\end{thebibliography}%

\end{document}